\documentclass[twocolumn,amsmath,amssymb,superscriptaddress]{revtex4}
\usepackage[utf8]{inputenc}
\usepackage{epsfig,amsmath,amssymb}
\usepackage{bm}
\usepackage{graphicx}
\usepackage{graphicx}
\usepackage{dcolumn,color}

\usepackage[colorlinks, linkcolor=red,anchorcolor=green,citecolor=blue]{hyperref}
\usepackage[normalem]{ulem}

\newcommand{\Tr}{\mathop{\mathrm{Tr}}}
\newcommand\diag{\operatorname{diag}}

\newcommand{\bx}{\bm{x}}
\newcommand{\bk}{\bm{k}}
\newcommand{\nn}{\nonumber}

\begin{document}

\title{Kubo formulae for first-order spin hydrodynamics}
\author{Jin Hu}
\affiliation{Department of Physics, Tsinghua University, Beijing 100084, China}
\begin{abstract}
We derive Kubo formulae for first-order spin hydrodynamics based on non-equilibrium statistical operators method. In first-order spin hydrodynamics, there are two new transport coefficients besides the ordinary ones appearing in first-order viscous hydrodynamics. They emerge due to the incorporation of the spin degree of freedom into fluids and the spin-orbital coupling. 
Zubarev's non-equilibrium statistical operator method can be well applied to investigate these quantum effects in fluids. The Kubo formulae, based on the method of non-equilibrium statistical operators, are related to equilibrium (imaginary-time) infrared Green's functions, and all the transport coefficients can be determined when the microscopic theory is specified.
  \end{abstract}

\maketitle

\section{Introduction}
Recent developments in relativistic heavy-ion collisions have seen great progress in studying observables with spin dependence. The measurements of spin polarization of $ \Lambda $ hyperons show that a fraction of the spin of quarks within the hyperons takes one particular direction \cite{STAR:2017ckg,Alpatov:2020iev}, which implies the media, quark-gluon plasma (QGP), should carry a large magnitude of angular momentum. Such a significant magnitude of vorticity leads to the phenomenon of spin alignments as a result of the well-known spin-orbital coupling. Theoretical researches on global polarization of $ \Lambda $ hyperons can be found in \cite{Wei:2018zfb,Karpenko:2016jyx,Csernai:2018yok,Li:2017slc,Bzdak:2017shg,Shi:2017wpk,Sun:2017xhx,Ivanov:2019wzg,Xie:2017upb}. The results of theoretical calculations fit the data well. Later, the STAR Collaboration published the measurements of differential spin polarization, namely, the dependence of $ \Lambda $ polarization  on the azimuthal angle and transverse momentum \cite{Adam:2019srw,Adam:2018ivw}. However, theoretical calculation can not provide satisfying explanation to experimental data, which is usually called ``spin sign problem'' \cite{Becattini:2017gcx,Xia:2018tes} , also see \cite{Liu:2020ymh} and \cite{Gao:2020vbh} for a review. To resolve this problem, new theoretical frameworks are necessary. One promising framework is hydrodynamics with the spin degree of freedom included. In other words, these direct experimental measurements of quantum effects in relativistic heavy-ion collisions motivate the incorporation of the quantum spin degree of freedom into the evolution of fluids.

To well describe the macroscopic dynamics of spin,
it is intuitive to generalize ordinary hydrodynamics, making it a spinful one.
There are many efforts following this direction. ``Ideal'' relativistic hydrodynamics with spin freedom was proposed in the context of the QGP~\cite{Florkowski:2017ruc}. Some  relevant discussions can also be seen in \cite{Becattini:2009wh,Montenegro:2017rbu,Florkowski:2018fap}. Recently, viscous spin hydrodynamics has also been put into consideration~\cite{Hattori:2019lfp,Fukushima:2020ucl}. In these works, two new transport coefficients arise reflecting new physical effects with spin freedom, which will be the main focus of our interest herein. 

Transport coefficients are important quantities manifesting the transport property of the medium in the field of heavy-ion collisions. In the case of spin hydrodynamics, new transport coefficients well capture the property of slow dynamics in the spinful fluid system \cite{Hattori:2019lfp}. The usual dissipative hydrodynamics have been widely used to describe the collective behavior of the QGP in the last decade. Once the spin degree of freedom was considered, as mentioned in \cite{Liu:2020ymh} , a numerical implementation of causal spin hydrodynamics is a promising one, particularly, it is hopeful to get insight into the ``spin sign problem''. For the purpose of quantitatively describing the evolution of the fluid system, transport coefficients are indispensable inputs. There are many methods for calculating the transport coefficients. The kinetic theory based on transport equation offers us an effective tool for investigating transport properties \cite{Xu:2008dv,Xu:2007ns}. Noting that transport methods, for example, Boltzmann equation, rely on the picture of quasi-particle, so we are supposed to apply them with caution. An alternative approach is based on the Kubo method, in which the correlation functions represent the response of an equilibrated system to a perturbation. Based on this method transport coefficients are directly linked to real-time retarded Green's functions, which can be evaluated by analytic continuation of equilibrium Green's functions formulated in imaginary time. 

In this paper, we utilize the non-equilibrium statistical operator method developed by
Zubarev~\cite{Zubarev,Hosoya:1983id,Horsley:1985dz} to derive Kubo formulae
for transport coefficients of relativistic spinful fluids. The non-equilibrium statistical operator is a generalization of the equilibrium Gibbs statistical operator to
non-equilibrium states. Using this approach, we are able to separate naturally the equilibrium part and the non-equilibrium part which takes the form of gradients of the thermodynamical parameters. From the viewpoint of linear response, transport coefficients can be obtained from linear perturbations of the non-equilibrium statistical operator around its equilibrium expectation value.

This paper is organized as follows. In Sec.~\ref{spin hydro}
we present a brief review of relativistic spin hydrodynamics in dissipative cases based on \cite{Hattori:2019lfp,Fukushima:2020ucl}. In Sec.~\ref{nonequilibrium} we adopt the non-equilibrium statistical operator method to derive Kubo formulae in first-order spin hydrodynamics,  which  relates transport coefficients to retarded correlation functions defined in terms of the underlying elementary fields. Discussion and outlook are given in Sec.~\ref{summary}. Natural units $\hbar=k_B=c=1$ are used. The metric tensor here is given by $g^{\mu\nu}=\diag(1,-1,-1,-1)$, while $\Delta^{\mu\nu} \equiv g^{\mu\nu}-u^\mu u^\nu$ is the projection tensor orthogonal to the four-vector fluid velocity $u^\mu$.
In addition, we employ the symmetric/antisymmetric shorthand notations:
\begin{eqnarray}
X^{( \mu\nu ) } &\equiv& (X^{ \mu\nu } + X^{ \nu \mu})/2, \\
X^{[ \mu\nu ] } &\equiv& (X^{ \mu\nu } - X^{ \nu \mu})/2, \\
X^{\langle \mu\nu \rangle}&\equiv&
	\bigg(\frac{\Delta^{\mu}_{\alpha} \Delta^{\nu}_{\beta} 
	+ \Delta^{\nu}_{\alpha} \Delta^{\mu}_{\beta}}{2}
	 - \frac{\Delta^{\mu\nu} \Delta_{\alpha\beta}}{3}\bigg)X^{\alpha\beta}.
\end{eqnarray}
\section{Review of first order spin hydrodynamics}\label{spin hydro}
Hydrodynamics is based on basic conservation laws \cite{Landau:Fluid}, which are conservation of the energy-momentum $T^{\mu\nu} $ and conserved current $N^{\mu}$ for the spinless case,
\begin{align}
\label{eq:Conserv-em}
&\partial_\mu T^{\mu\nu} =0 \, ,
\\
\label{eq:Conserv-J}
&\partial_\mu N^{\mu} = 0 \, .
\end{align}
Problem comes when we need to take into account the spin degree of freedom.  Spin angular momentum  plays a big role in the evolution of spinful fluids, and one needs to refer to another conserve law: the conservation of total angular momentum, which is expressed as:
\begin{align}
\label{eq:angular}
\partial_\lambda \Sigma^{\lambda\mu\nu} =0 
\, .
\end{align}
Microscopically, the rank three tensor $\Sigma^{\lambda\mu\nu}$ can be decomposed into two distinct components by calculating Noether current 
for the Lorentz symmetry which reads:
$ \Sigma^{\mu\alpha\beta}
= (x^\alpha T^{\mu\beta}  - x^\beta T^{\mu\alpha} ) +  S^{\mu\alpha\beta} $, where $ S^{\mu\alpha\beta} = - S^{\mu\beta\alpha} $ \cite{Fukushima:2020qta}.
$ S^{\mu\alpha\beta}  $ arises from the invariance with respect to
the representation of the Lorentz group acting on a field under consideration,
and is naturally identified with spin angular momentum.
On the other hand, the orbital angular momentum part comes
from the coordinate transformation of the argument of the field.
Here $ T^{\mu\nu}$ is  the {\it canonical} energy-momentum tensor, featured as  having both symmetric and antisymmetric
components: $ T^{\mu\nu} \equiv T^{\mu\nu}_{(s)} + T^{\mu\nu}_{(a)} $.   In this work, we will keep using canonical form of energy-momentum tensor. Discussions about details of pseudo gauge transformed form, for instance, {\it Belinfante} form can be seen in \cite{Fukushima:2020ucl}.
Explicitly (\ref{eq:angular}) can be rewritten as:
\begin{align}
\label{eq:angular1}
\partial_\lambda S^{\lambda\mu\nu} = T^{\nu\mu}-T^{\mu\nu}
\, .
\end{align}

First, recalls that the thermodynamic relation in equilibrium as well as the first law of thermodynamics, which read as:
\begin{eqnarray}
\label{eq:FirstLaw}
&&
Ts+\mu n = e+p - \omega_{\mu\nu} S^{\mu\nu}, \\
&&  T ds + \mu dn = de - \omega_{\mu\nu} d S^{\mu\nu},
\\
&&
s dT + nd\mu = dp - S^{\mu\nu} d\omega_{\mu\nu} ,
\end{eqnarray}
where $ T $, $ s $, $ \mu $, $ n $, $e  $, and $p  $ denote
the local temperature, entropy density, chemical potential, conserved charge density, energy density, and pressure, respectively.
In this paper, we consider only one conserved charge.
Here, analogous to the relation of chemical potential and charge density,  a ``spin potential'' $\omega_{\mu\nu} $ is introduced conjugate to the spin density $ S^{\mu\nu} $. And one thing that needs to be paid attention to is $\omega_{\mu\nu} = O(\partial^1)$ in derivative expansion, and we will show the reason for this counting later.

On the basis of a derivative expansion, we obtain the constitutive relations:
\label{eq:constitutive}
\begin{align}
&
T^{\mu\nu} = e u^\mu u^\nu + p \Delta^{\mu\nu} + T^{\mu\nu}_{(1)}
\, ,
\\
	&
	N^{\mu} = n u^\mu + j^{\mu}_{(1)}
	\, ,
\\
&
\Sigma^{\mu\alpha\beta} =  u^\mu S^{\alpha\beta} + \Sigma^{\mu\alpha\beta} _{(1)}
\, .
\end{align}
The normalization of the fluid velocity 
reads $u^\mu u_\mu = 1$, and we also use $\nabla^\mu \equiv \Delta^{\mu\nu}\partial_\nu$, $D\equiv u^\mu\partial_\mu$ as the spatial and temporal  component of derivative, respectively. It is not hard to notice that the spin density $S^{\mu\nu}$ satisfies the antisymmetric property $S^{\mu\nu} =- S^{\nu\mu}$. Accordingly, we have $\omega_{\mu\nu} = - \omega_{\nu\mu}$.
The thermodynamic second law puts additional limits onto the entropy production. Following the prescription of \cite{Israel:1979wp}, we make assumptions about $s^\mu$ in the presence of spin freedom:
\begin{align}
\label{eqs1}
s^\mu
& = \frac{u_\nu}{T}T^{\mu\nu} +\frac{p}{T}u^{\mu} s -\frac{\mu}{T}j^{\mu} -\frac{1}{T}\omega_{\alpha\beta}S^{\alpha\beta}u^{\mu}+O(\partial^2) 
\nn
\\
&=su^\mu + \frac{u_\nu}{T}T^{\mu\nu}_{(1)}-\frac{\mu}{T}j^{\mu}_{(1)} + O(\partial^2) .
\end{align}
Combined with thermodynamic relation (\ref{eq:FirstLaw}) and the constituent equation of energy, $u_\nu\partial_\mu T^{\mu\nu}=0$ , the entropy production is simplified as:
\begin{equation}
\label{eqs2}
\partial_\mu s^\mu = -j^{\mu}_{(1)}\partial_\mu\frac{\mu}{T}+T^{\mu\nu}_{(1)}\partial_\mu\frac{u_\nu}{T}+\frac{1}{T}\omega_{\alpha\beta}\partial_\mu(S^{\alpha\beta}u^{\mu})
\, .
\end{equation}
Taking into account that $T^{\mu\nu}_{(1)}$ has the symmetric and antisymmetric part, we then have:
\begin{eqnarray}
&&
\label{eq:TUV(1)}
T^{\mu\nu}_{(1s)} = 2h^{(\mu}u^{\nu)}+\pi^{\mu\nu}+\Pi\Delta^{\mu\nu}, \\
&& T^{\mu\nu}_{(1a)} = 2q^{[\mu}u^{\nu]}+\tau^{\mu\nu} ,
\end{eqnarray}
where $\pi^{\mu\nu}$ and $\Pi$ represent shear stress tensor and bulk viscous pressure, and $h^\mu$ is heat flow. Meanwhile, $\tau^{\mu\nu}$ and $q^\mu$ are antisymmetric counterparts of $\pi^{\mu\nu}$ and $h^\mu$, respectively. These five quantities are all of the first order in gradient expansion. One can further find $ \pi^{\mu\nu} = \pi^{\nu\mu} $, $\tau^{\mu\nu} =-\tau^{\nu\mu} $,
and $ h^\mu u_\mu = q^\mu u_\mu= \tau^{\mu\nu} u_\nu = \pi^{\mu\nu}u_\nu= 0 $.
Putting $T^{\mu\nu}_{(1)}$ into $\partial_\mu s^\mu$, and neglecting the terms of $O(\partial^3)$, we obtain full form of the entropy production in first-order spin hydrodynamics: 
\begin{align}
\label{eq:fullsu}
\partial_\mu s^\mu
& = \Big(h^\mu-\frac{e+p}{n}j^{\mu}_{(1)}\Big)\frac{n}{e+p}\nabla_\mu \frac{\mu}{T} 
	+\frac{\pi^{\mu\nu}}{T}\partial_{\langle\mu} u_{\nu\rangle}
\nn
\\
&\
-\frac{\Pi}{T}\theta + q^\mu \Big(-\frac{u\cdot \partial}{T}u_\mu+\partial_\mu\frac{1}{T}+\frac{4\omega_{\mu\nu}u^\nu}{T}\Big)
\nn
\\
&\
+\tau^{\mu\nu} \Big[\Delta_{\mu\rho}\Delta_{\nu\sigma}\Big(\partial^\rho \frac{u^\sigma}{T}-\partial^\sigma \frac{u^\rho}{T}\Big)+2\frac{\omega_{\mu\nu}}{T}\Big],
\end{align}
where the notation $\theta=\partial_\mu u^\mu$ is used. Noting that, when the system is in equilibrium, the entropy production must cease and we obtain $\omega_{\mu\nu}=-\frac{T}{2}\omega^{th}_{\mu\nu}$ with the thermal vorticity $\omega^{th}_{\mu\nu}=\Delta_{\mu\rho}\Delta_{\nu\sigma}(\partial^\rho \frac{u^\sigma}{T}-\partial^\sigma \frac{u^\rho}{T})$~\cite{Becattini:2012tc,Becattini:2018duy}. According to this  argument, the counting of $\omega_{\mu\nu}$ can be estimated as $O(\partial^1)$ assuming the system is not far away from global equilibrium. Following the routine of first-order hydrodynamics, we impose the sufficient conditions of semipositive entropy production $\partial_\mu s^\mu\geq0$, this is, cast every term into positive semidefinite quadratic form so that the entropy production can be seen as a sum of squares. Therefore we have:
\begin{eqnarray}
&&
\label{shear}
\pi^{\mu\nu}=2\eta\nabla^{\langle\mu}u^{\nu\rangle}, \\
\label{bulk}
&&  \Pi=-\zeta\theta,
\\
&&
\label{heat}
h^\mu-\frac{e+p}{n}j^\mu_{(1)}=-\kappa\frac{nT}{e+p}\nabla^\mu\frac{\mu}{T} ,
\\
&&
\label{tau}
\tau^{\mu\nu}=2\gamma \big(\omega_{th}^{\mu\nu}+2\Delta^{\mu\rho}\Delta^{\nu\sigma}\omega_{\rho\sigma} \big),
\\
&&
\label{qmu}
q^\mu=\lambda \big(Du^\mu+\frac{\nabla^\mu T}{T}-4\omega^{\mu\nu}u_\nu \big),
\end{eqnarray}
$\eta$, $\zeta$ and $\kappa$ represent shear viscosity, bulk viscosity and heat conductivity respectively,  $\gamma$ and $\lambda$ are new transport coefficients of spin hydrodynamics, which are identified as ``rotational viscosity'' in ~\cite{degroot} and ``boost heat conductivity'' in \cite{Hattori:2019lfp}. In the next section, we will derive Kubo formulae for these transport coefficients.

\section{ Non-equilibrium Statistical Operators and Kubo formulae} \label{nonequilibrium}
The method of non-equilibrium statistical operators
(NESO) developed by Zubarev starts from constructing a statistical ensemble encoding thermodynamic information of the macroscopic state of the system in non-equilibriumn state.
In present case we consider the system in the
hydrodynamic regime which is near local equilibrium and thermodynamic parameters such as temperature and chemical potentials can be well defined locally. See \cite{Hosoya:1983id} and \cite{Huang:2011dc} for reference about NESO. Following Zubarev's practice, the form of NESO in the textbook is \cite{Zubarev,Hosoya:1983id,Huang:2011dc} 
\begin{eqnarray}
&\hat{\rho}(t)=Q^{-1}\exp\left[-\int d^3\bx\, \hat{Z}(\bx,t)\right],\\
&Q=\Tr\exp\left[-\int d^3\bx\, \hat{Z}(\bx,t)\right],
\end{eqnarray}
where the operator $\hat{Z}$ is defined as
\begin{align} 
\label{Z}
&\hat{Z}(\bx,t)=\epsilon\int_{-\infty}^t dt^\prime e^{\epsilon(t^\prime-t)}
\Big[\beta^\nu(\bx,t^\prime) \hat{T}_{0\nu}(\bx,t^\prime)
\nn
\\
&\ \ \qquad-\alpha(\bx,t^\prime)\hat{N}^0(\bx,t^\prime)-\frac{1}{2}\omega_{\rho\sigma}(\bx,t^\prime)\hat{S}^{0\rho\sigma}(\bx,t^\prime) \Big],
\end{align}
with $\epsilon\rightarrow+0$ after taking thermodynamic limit. Here we have introduced new Lagrange multiplier $\omega^{\rho\sigma}(\bx,t)$ and the operator $\hat{S}^{0\rho\sigma}$ coupled to it, which can be understood as the incorporation of total angular momentum. From the form of total angular momentum $\Sigma^{\mu\alpha\beta}$, we can deduce that the conservation condition of total angular momentum brings only new information in the part of spin angular momentum (the information of orbital part can be reproduced by energy-momentum tensor).
More details can be found in \cite{Becattini:2018duy}. Other Lagrange multipliers are written explicitly as:
\begin{eqnarray}
\label{beta}
&\beta^\nu(\bx,t)=\beta(\bx,t) u^\nu(\bx,t),\\ 
\label{alpha}
&\alpha(\bx,t)=\beta(\bx,t)
\mu(\bx,t).
\end{eqnarray}
The parameters $\beta$  stand for the
inverse local equilibrium temperature. We need to identify these parameters in the language of statistical operators, which will be deferred below.
We here express three local conservation laws with statistical operators:
\begin{eqnarray}
\label{conserve} \partial_\mu\hat{T}^{\mu\nu}=0\;,\;\; \partial_\mu\hat{N}^{\mu}=0\;,\;\;
\partial_\mu\hat{S}^{\mu\rho\sigma}=\hat{T}^{\sigma\rho}-\hat{T}^{\rho\sigma}.
\end{eqnarray}
Integrating Eq.(\ref{Z}) by parts and utilizing Eq.(\ref{conserve}), we can get:
\begin{align}
\label{Zint} 
&\int d^3\bx\,\hat{Z}(\bx,t) = \hat{A}-\hat{B} ,
\\
\label{A}
&\hat{A}=\int d^3\bx \Big[\beta^\nu(\bx,t)
\hat{T}_{0\nu}(\bx,t)-\alpha(\bx,t)\hat{N}^0(\bx,t)
\nn
\\
&\quad
-\frac{1}{2}\omega_{\rho\sigma}(\bx,t)\hat{S}^{0\rho\sigma}(\bx,t)\Big],
\\
\label{B}
&
\hat{B}=\int d^3\bx\int_{-\infty}^t dt^\prime e^{\epsilon(t^\prime-t)}
\Big[\hat{T}_{\mu\nu}(\bx,t^\prime)\partial^\mu\beta^\nu(\bx,t^\prime)
\nn
\\
&\quad
-\hat{N}^\mu(\bx,t^\prime)\partial_\mu\alpha(\bx,t^\prime)
+\omega_{\mu\nu}(\bx,t^\prime)\hat{T}^{[\mu\nu]}(\bx,t^\prime)\Big].
\end{align}
In deriving Eq.(\ref{Zint}), we notice that integrating by parts will bring in three-dimensional surface integrals that are often discarded, but for temporal dimension, it is different because of definite integration upper limit. If we take $t$ to infinity, the derivative term would all vanish showing that the system should go to equilibrium given long enough evolution time. Due to the counting of $\omega_{\mu\nu}$, the term of $\partial_\mu\omega^{\mu\nu}$ is the order of $O(\partial^2)$, so we neglect this term in Eq.(\ref{Zint}). Following the spirit of non-equilibrium statistical mechanics, we treat the derivative terms as thermodynamic forces that lead to dissipation. By doing so, we are able to decompose statistical operators into the local equilibrium part and non-equilibrium part. We define the local equilibrium statistical operator as:
\begin{align}
\label{leso} 
&\hat{\rho}_{\rm{leq}}\equiv Q_{\rm{leq}}^{-1}\exp\Big(-\hat{A}\,\Big),
\\
&Q_{\rm{leq}}=\Tr\exp\Big(-\hat{A}\,\Big),
\end{align}
and the complete statistical operator as:
\begin{align}
\label{feso} 
&\hat{\rho}\equiv Q^{-1}\exp\Big(-\hat{A}+\hat{B}\,\Big),\\
&Q=\Tr\exp\Big(-\hat{A}+\hat{B}\,\Big).
\end{align}
Now comes the question of how to handle the complete statistical operator taking in the form of the exponential function of the sum of two operators. Noting $[\hat{A},\hat{B}]\neq0$ , $[\hat{A},[\hat{A},\hat{B}]]\neq0$ and $[\hat{B},[\hat{A},\hat{B}]]\neq0$, the form of exponential expansion is complex. Here we adopt an approach of operators expansion proposed in \cite{Zubarev}. 
We focus on small perturbation around the equilibrium system, that is, the thermodynamic forces can be treated as perturbations. In this case, it is safe to say that the relation of these forces and irreversible dissipative currents is linear so that we can expand Eq.(\ref{feso}), keep only linear term, and approximate the complete statistical operator as:
\begin{eqnarray}
\label{neso3} \hat{\rho}=\left[1+\int_0^1 d\tau \left( e^{-\hat{A}\tau}\hat{B}
e^{\hat{A}\tau}-\langle\hat{B}\rangle_{\rm{leq}} \right) \right]\hat{\rho}_{\rm{leq}},
\end{eqnarray}
where $\langle\hat{B}\rangle_{\rm{leq}}=\Tr[\hat{\rho}_{\rm{leq}}\hat{B}]$ is the expectation over the local equilibrium operator. We consider energy-momentum tensor in non-equilibrium state. First, we evaluate the energy-momentum tensor averaged over the non-equilibrium distribution:
\begin{align}
\label{Tuv} 
&\langle{\hat{T}^{\mu\nu}(\bx,t)}\rangle=\langle{\hat{T}^{\mu\nu}(\bx,t)}\rangle_{\rm{leq}}+\int d^3\bx^\prime\int_{-\infty}^t dt^\prime e^{\epsilon(t^\prime-t)}
\nn
\\
&\qquad \quad \quad \ \ 
\Big[\Big(\hat{T}^{\mu\nu}(\bx,t) \,,\, \hat{T}^{\rho\sigma}(\bx^\prime,t^\prime)\Big) 
	\partial_\rho\beta_\sigma(\bx^\prime,t^\prime)
\nn
\\
&\qquad \quad \quad \ \ 
-\Big(\hat{T}^{\mu\nu}(\bx,t) \,,\, \hat{N}^{\rho}(\bx^\prime,t^\prime)\Big) 
\partial_\rho\alpha(\bx^\prime,t^\prime)
\nn
\\
&\qquad \quad \quad \ \
+\Big( \hat{T}^{\mu\nu}(\bx,t) \,,\, \hat{T}^{[\rho\sigma]}(\bx^\prime,t^\prime)\Big)
	\omega_{\rho\sigma}(\bx^\prime,t^\prime)\Big],
\end{align}
where $\langle\hat{B}\rangle=\Tr[\hat{\rho}\hat{B}]$ is the expectation over the
complete operator, with the binary operator
\begin{align}
\label{correlator}
&\left(\hat{X}(\bx,t),\hat{Y}(\bx',t')\right) \nn \\
&\equiv\int_0^1 d\tau\Big\langle\hat{X}(\bx,t)\big( e^{-\hat{A}\tau}\hat {Y}(\bx',t')e^{\hat{A}\tau}
-\langle\hat{Y}(\bx',t')\rangle_{\rm{leq}}\big)\Big\rangle_{\rm{leq}}
\end{align}
being the Kubo correlation function.

We precede to match the relevant tensors $T^{\mu\nu}$ and $N^{\mu}$  in hydrodynamics with the corresponding operators. A straightforward and natural tensor decomposition reads as:
\begin{align}
\label{hatT}
&
\hat{T}^{\mu\nu} = \hat{e} u^\mu u^\nu - \hat{p} \Delta^{\mu\nu} + \hat{T}^{\mu\nu}_{(1s)} +\hat{T}^{\mu\nu}_{(1a)} 
\, ,
\\
\label{hatTs}
&
\hat{T}^{\mu\nu}_{(1s)} = 2\hat{h}^{(\mu}u^{\nu)}+\hat{\pi}^{\mu\nu}+\hat{\Pi}\Delta^{\mu\nu} \, ,
\\
\label{hatTa}
&
\hat{T}^{\mu\nu}_{(1a)} = 2\hat{q}^{[\mu}u^{\nu]}+\hat{\tau}^{\mu\nu},
\, \\
&\hat{N}^{\mu}=\hat{n}u^\mu+\hat{j}^\mu_{(1)},
\end{align}
which consistently  matches the form of ${T}^{\mu\nu}$ and $N^\mu$ in hydrodynamics. We also need to specify the parameters coupled to $\hat{T}^{\mu\nu}$, $\hat{N}^{\mu}$ and $\hat{S}^{\lambda\mu\nu}$. By imposing Landau matching conditions 
$u^\mu\delta\langle\hat{T}_{\mu\nu}\rangle u^\nu
=0$ , $u^\mu\delta\langle\hat{N}_{\mu}\rangle=0$ \cite{Landau:Fluid} with the notation $\delta\langle\hat{X}\rangle=\langle\hat{X}\rangle-\langle\hat{X}\rangle_{\rm{leq}}$, the parameters $\beta$ and $\alpha$ are identified as the inverse of temperature and the ratio of chemical potential to temperature. As for the identification of $\omega_{\mu\nu}$ with spin potential, details can be found in \cite{Becattini:2018duy} and we will keep using this identification hereafter. Then the expectation of $\hat{T}_{\mu\nu}$ is evaluated over non-equilibrium operator and compared with the result with that of first-order spin hydrodynamics. To that end, we first rewrite the terms $\hat{T}^{\rho\sigma}\partial_\rho\beta_\sigma$, $N^\mu\partial_\mu \alpha$ and $\hat{T}^{\rho\sigma}\omega_{\rho\sigma}$ with the identification of these Lagrange multipliers:
\begin{align}
\label{tempory}
&\hat{T}^{\mu\nu}\partial_\mu\beta_\nu = \beta\hat{\pi}^{\mu\nu}\partial_\mu u_\nu+\beta
\hat{h}^\mu\big(\beta^{-1}\partial_\mu\beta+Du_\mu\big)
\nn
\\
&\qquad \qquad
+\beta\hat{\tau}^{\mu\nu}\partial_\mu u_\nu+\beta\hat{q}^\mu\big(-\beta^{-1}\partial_\mu\beta+Du_\mu\big)
\nn
\\
&\qquad \qquad
+\hat{e}D\beta-\beta\big(\hat{p}-\hat{\Pi}\big)\theta ,\\ 
\label{tempory1}
&\hat{N}^{\mu}\partial_\mu\alpha =\hat{n}D\alpha+\hat{j}^\mu_{(1)}\nabla_\mu\alpha,\\
&\hat{T}^{[\mu\nu]}\omega_{\mu\nu} = 2\hat{q}^{[\mu}u^{\nu]}\omega_{\mu\nu}+\hat{\tau}^{\mu\nu}\omega_{\mu\nu}.
\end{align}
In order to match thermodynamic forces in first-order hydrodynamics, we substitute $D\beta$ and $D\alpha$ with $\theta$ by using the equations of zero-order hydrodynamics,
\begin{align}
&\partial_\mu \langle \hat{T}^{\mu\nu}\rangle_{\rm{leq}}=0,\\
&\partial_\mu \langle \hat{N}^{\mu}\rangle_{\rm{leq}}=0.
\end{align}
Taking the scalar product of these equations with the four velocity $u^\nu$, we get:
\begin{align}
&D\langle \hat{e}\rangle_{\rm{leq}}=-(\langle\hat{e}\rangle_{\rm{leq}}+\langle \hat{p}\rangle_{\rm{leq}})\theta,\\
&D\langle\hat{n}\rangle_{\rm{leq}}=-\langle\hat{n}\rangle_{\rm{leq}}\theta.
\end{align}
Noting that the matching conditions ensures that $e=\langle\hat{e}\rangle_{\rm{leq}}$ and $n=\langle\hat{n}\rangle_{\rm{leq}}$, we shall using these notations in the following paragraphs. Straightforward calculation leads us to:
\begin{align}
&D\beta = \frac{\theta}{J}\Big(-\big(e+p\big)\frac{\partial n}{\partial\alpha}+n\frac{\partial e}{\partial\alpha}\,\Big),\\
&D\alpha = \frac{\theta}{J^\prime}\Big(-\big(e+p\big)\frac{\partial n}{\partial\beta}+n\frac{\partial e}{\partial\beta}\,\Big),\\
J=&\frac{\partial e}{\partial\beta}\frac{\partial n}{\partial\alpha}-\frac{\partial n}{\partial\beta}\frac{\partial e}{\partial\alpha},\quad J^\prime=\frac{\partial e}{\partial\alpha}\frac{\partial n}{\partial\beta}-\frac{\partial n}{\partial\alpha}\frac{\partial e}{\partial\beta},
\end{align}
with the derivative of the thermodynamic functions with respect to $\alpha$ calculated holding $\beta$ fixed and vice versa.

Now Eq.({\ref{tempory}}) and ({\ref{tempory1}}) can be cast into:
\begin{align}
\label{tuvbeta}
&\hat{T}^{\mu\nu}\partial_\mu\beta_\nu \nn = \beta\hat{\pi}^{\mu\nu}\partial_\mu u_\nu+\beta
\hat{h}^\mu\left(\beta^{-1}\partial_\mu\beta+Du_\mu\right)
\nn
\\
&\qquad \qquad 
+\beta\hat{\tau}^{\mu\nu}\partial_\mu u_\nu+\beta\hat{q}^\mu\left(-\beta^{-1}\partial_\mu\beta+Du_\mu\right)
\nn
\\
&\qquad \qquad
-\beta\hat{p}^\prime\theta , \\ 
&\hat{N}^{\mu}\partial_\mu\alpha=\hat{j}^\mu_{(1)}\nabla_\mu\alpha+\frac{\hat{n}}{J^\prime}\Big[-\big(e+p\big)\frac{\partial n}{\partial\beta}+n\frac{\partial e}{\partial\beta}\,\Big]\theta,\\
\label{pprime}
& 
\hat{p}^\prime=\hat{p}-\hat{\Pi}-\frac{1}{J\beta}\Big[-\big(e+p\big)\frac{\partial n}{\partial\alpha}+\,n\frac{\partial e}{\partial\alpha}\,\Big]\,\hat{e}.
\end{align}
From now on we will handle Kubo correlations so as to get final results.
In an isotropic medium, one can turn to Curie's principle for help and that has been used  to simplify Eq.({\ref{Tuv}}). Curie's principle shows that the correlation function between operators of different ranks and spatial parity vanishes. The remaining Kubo correlations can be expressed in the comoving frame as:
\begin{align}
\label{correlators}
& \left(\hat{h}_{k}^\prime,\hat{h}_{l}^\prime \right)=L_{h^\prime}\delta_{kl} ,\nn \\
& \left(\hat{\pi}^{kl},\hat{\pi}^{mn}\right)=L_\pi\frac{1}{2}\left(\delta^{km}\delta^{ln}+\delta^{kn}\delta^{lm}-\frac{2}{3}\delta^{kl}\delta^{mn}\right) ,\nn \\
&
\left(\hat{q}^k,\hat{q}^l\right)=L_q\delta^{kl} ,\nn \\
& \left(\hat{\tau}^{kl},\hat{\tau}^{mn}\right)=L_\tau\frac{1}{2}\left(\delta^{km}\delta^{ln}-\delta^{kn}\delta^{lm}\right) ,
\end{align}
with $\hat{h}^\prime_\mu=\hat{h}_\mu-\frac{e+p}{n}\hat{j}_\mu$, $\delta^{ij}$ being the Kronecker symbol, and $L_i$ are scalar functions that can be determined by taking trace in both sides of Eq.(\ref{correlators}). And we conclude that the correlation between symmetric tensor and antisymmetric tensor is zero and we needn't consider the contribution of these cross parts. A simple explanation is put following. First, we assume that:
\begin{align} 
\label{ansatz}
\left(\hat{\pi}^{kl},\hat{\tau}^{mn}\right)\partial_m u_n=L_1\partial^ku^l+L_2\partial^lu^k+L_3\delta^{kl}\theta,  
\end{align} 
with $L_i(i=1,2,3)$ being scalar functions. The reason why we can make such an ansatz is that we have constrained the form of current-force relation as Eq.(\ref{shear}). It is this simple linear current-force relation that leads to this ansatz. Moving on, for the symmetry of exchanging $\mu$ and $\nu$, $L_1$ must equal $L_2$. Equivalently, we have:
\begin{align}
\label{ansatz1}
\left(\hat{\pi}^{kl},\hat{\tau}^{mn}\right)=L_1\left(\delta^{km}\delta^{ln}+\delta^{kn}\delta^{lm}\right)+L_3\delta^{kl}\delta^{mn}.
\end{align}
Because the right-hand side of Eq.(\ref{ansatz1}) is not antisymmetric when exchanging $m$ and $n$, which is in conflict with the left-hand side, the correlation between symmetric tensor and antisymmetric tensor is exactly zero. Then we boost to general reference frame, substitute the Kronecker symbol $\delta^{\mu\nu}$ with $-\Delta^{\mu\nu}$, and take trace to get all $L$ functions \cite{Zubarev,Hosoya:1983id}.

In order to extract transport coefficients, we suppose the changes of thermodynamic forces within correlation length are sufficiently small so that we can factorize them out of Eq.(\ref{Tuv}). Thus we obtain the linear thermodynamic current-force relation combining Eqs.(\ref{hatT}), (\ref{hatTs}), (\ref{hatTa}), (\ref{tuvbeta}), (\ref{pprime}), and (\ref{correlators}) together:
\begin{eqnarray}
&&
\label{piuv}
\langle\hat{\pi}^{\mu\nu}\rangle=2\eta\nabla^{\langle\mu}u^{\nu\rangle}, \\
&&  \langle\hat{\Pi}\rangle=-\zeta\theta,
\\
&&
\label{hu}
\langle\hat{h}^\mu\rangle - \frac{e+p}{n}\langle\hat{j}_{(1)}^\mu\rangle  = -\kappa \frac{nT}{e+p}\nabla^\mu\frac{\mu}{T},
\\
&&
\langle\hat{\tau}^{\mu\nu}\rangle=2\gamma\Big(\omega_{th}^{\mu\nu}+2\Delta^{\mu\rho}\Delta^{\nu\sigma}\omega_{\rho\sigma}\Big), 
\\
&&
\langle\hat{q}^\mu\rangle=\lambda\bigg(Du^\mu+\frac{\nabla^\mu T}{T}-4\omega^{\mu\nu}u_\nu\bigg) ,
\end{eqnarray}
where the Gibbs--Duhem relation has been employed. 
By comparing these equations with Eqs.(\ref{shear}), (\ref{bulk}), (\ref{heat}), (\ref{tau}) and (\ref{qmu}), we conclude that we have reproduced the linear law of first-order spin hydrodynamics with the method of statistical operators. Throughout the derivation, the four-vector fluid velocity is not specified, which means the results we obtain is frame independent up to first order in gradients. After finishing the above manipulations, all the transport coefficients can be given in terms of Kubo correlation functions:
\begin{align}
\label{eta} 
\eta=&\frac{\beta}{10}\int d^3\bx^{\prime}\int_{-\infty}^t dt^{\prime}
e^{\epsilon(t^{\prime}-t)}\Big( \hat{\pi}^{\mu\nu}(\bx,t), \hat{\pi}_{\mu\nu}(\bx^{\prime},t^{\prime})\Big),\\
\label{kappa} 
\kappa=&\frac{-\beta}{3}\int d^3\bx^{\prime}\int_{-\infty}^t dt^{\prime}
e^{\epsilon(t^{\prime}-t)}\Big( \hat{h}^{\prime\mu}(\bx,t), \hat{h}_{\mu}^\prime(\bx^{\prime},t^{\prime})\Big),\\
\label{zeta} 
\zeta=&\beta\int d^3\bx^{\prime}\int_{-\infty}^t dt^{\prime} e^{\epsilon(t^{\prime}-t)}
	\Big( \hat{p}^\star(\bx,t), \hat{p}^\star(\bx^\prime,t^\prime)\Big),\\
\label{gamma} 
\gamma=&\frac{-\beta}{6}\int d^3\bx^\prime\int_{-\infty}^t dt^\prime
e^{\epsilon(t^\prime-t)}\Big( \hat{\tau}^{\mu\nu}(\bx,t), \hat{\tau}_{\mu\nu}(\bx^\prime,t^\prime)\Big),\\
\label{lambda} 
\lambda=&\frac{-\beta}{3}\int d^3\bx^{\prime}\int_{-\infty}^t dt^{\prime}
e^{\epsilon(t^{\prime}-t)}\Big(\hat{q}^\mu(\bx,t) , \hat{q}_\mu(\bx^{\prime},t^{\prime})\Big),
\end{align}
with $\hat{p}^*=\hat{p}^\prime+\frac{\hat{n}}{\beta\theta}D\alpha$.

In the following paragraphs, we will build up the connection between Kubo correlation functions and retarded Green functions. The discussion of the following paragraphs is similar to that in \cite{Huang:2011dc}. Direct evaluation of Eq.(\ref{correlator}) leads to:
\begin{align}
\label{xycorrelate}
&\left(\hat{X}(\bx,t),\hat{Y}(\bx',t')\right)
\nn
\\
&\equiv\int_0^1 d\tau\Big\langle
\hat{X}(\bx,t)\left[ e^{-\hat{A}\tau}\hat{Y}(\bx',t')e^{\hat{A}\tau}-\langle{\hat
	Y}(\bx',t')\rangle_{\rm{leq}}\right]\Big\rangle_{\rm{leq}} 
\nn
\\
&
=\frac{i}{\beta}\int_{-\infty}^{t'}ds \Big\langle
\Big[{\hat X}(\bx,t), {\hat Y}(\bx',s)\Big]\Big\rangle_{\rm{leq}},
\end{align}
where we supposed in the last step that the correlation of two operators vanishes when time goes to distant past. In deriving Eq.(\ref{xycorrelate}), we have utilized the conclusion that $\hat{A}$ can be treated as Hamiltonian operator. To see that, we are informed of three points. First, when we choose local rest frame or comoving frame, the first term within the integrand of Eq.(\ref{A}) is exactly $\beta\hat{H}$. Second, taking the second term means we are doing the calculation of the grand canonical ensemble. In finite temperature field theory, this term can always be added to Hamiltonian to construct the grand canonical Hamiltonian operator. Third, the third term related to the coupling of spin and vorticity, which is the covariant form of the scalar product of angular velocity and angular momentum $\bm{\omega}\cdot\bm{J}$ \cite{Becattini:2012tc}, can also be included in Hamiltonian in a rotating system. Combining these three considerations, we conclude that $e^{\hat{A}\tau}$ is a quantum mechanical evolution operator. Keep following this procedure:
\begin{align}
\rm{I}&=\int d^3\bx'\int_{-\infty}^t dt'
e^{\epsilon(t'-t)}\frac{i}{\beta}\int_{-\infty}^{t'}ds \Big\langle \Big[{\hat
	X}(\bx,t), {\hat Y}(\bx',s)\Big]\Big\rangle_{\rm{leq}}
\nn
\\
&
=-\int d^3\bx'\int_{-\infty}^t dt'e^{\epsilon(t'-t)}\frac{1}{\beta}\int_{-\infty}^{t'}ds
G_R(\bx-\bx',t-s)
\nn
\\
&
=\frac{i}{\beta}\lim_{\omega\rightarrow 0}
\lim_{\bk\rightarrow 0}\frac{\partial}{\partial\omega}G_R(\bk,\omega).
\end{align}
In obtaining this equation, the definition of retarded Green function is required:
\begin{align} G_{\hat{A}\hat{B}}^R(\bx,t)\equiv-i\theta(t)\left[\hat{A}(\bx,t),\hat{B}({\bf0},0)\right].
\end{align}
 So far we have proved Kubo correlation is exactly related to retarded Green function. Because formulae for transport coefficients are all relevant to self-correlation, we keep our focus on this case. Suppose $A,B$ represent the same operator, the imaginary(real) part of retarded Green is even(odd) function of $\omega$ according to the Onsager’s reciprocal principle \cite{Zubarev} such that:
\begin{align}
\rm{I}=-\frac{1}{\beta}\lim_{\omega\rightarrow 0}
\lim_{\bk\rightarrow 0}\frac{\partial}{\partial\omega}ImG^R_{\hat{A}\hat{A}}(\bk,\omega).
\end{align}

Collect all the results we obtained:
\begin{align}
\label{restkubo} 
&\eta=-\frac{1}{10}\lim_{\omega\rightarrow 0}
\frac{\partial}{\partial\omega}\rm{Im} G^R_{\hat{\pi}\hat{\pi}}(\b0,\omega),
\\
&
\zeta=-\lim_{\omega\rightarrow 0}
\frac{\partial}{\partial\omega}\rm{Im} G^R_{\hat{p}*\hat{p}*}(\b0,\omega),
\\
&
\kappa=\frac{1}{3}\lim_{\omega\rightarrow 0}
\frac{\partial}{\partial\omega}\rm{Im} G^R_{\hat{h}^\prime\hat{h}^\prime}(\b0,\omega),
\\
&
\gamma=\frac{1}{6}\lim_{\omega\rightarrow 0}
\frac{\partial}{\partial\omega}\rm{Im} G^R_{\hat{\tau}\hat{\tau}}(\b0,\omega),
\\
&
\lambda=\frac{1}{3}\lim_{\omega\rightarrow 0}
\frac{\partial}{\partial\omega}\rm{Im} G^R_{\hat{q}\hat{q}}(\b0,\omega).
\end{align}
The operators arising in the subscripts are all defined in the previous paragraphs. And the first three transport coefficients are consistent with the results of \cite{Huang:2011dc} and \cite{Hosoya:1983id}. We note that there is a factor $2$ difference compared to the result of $\eta$ in \cite{Hosoya:1983id}, which is due to the different definition of shear viscosity, [see Eq.(\ref{shear})]. The last two transport coefficients are exactly what we want, which can give a description of new transport properties of spinful fluids.
\\[10pt]

\section{Summary and Outlook} \label{summary}

We have evaluated Kubo formulae for transport coefficients arising in first-order spin hydrodynamics based on the approach of the non-equilibrium statistical operator. 
We apply Zubarev's statistical operator method to linearize the non-equilibrium corrections, and study how a thermal system respond to such linear perturbations. The Kubo formulae we obtained are related to equilibrium (imaginary-time) infrared Green's functions.
Given specific microscopic theory, the imaginary-time Green's functions in finite temperature field can be formulated. Thus, by analytical continuation, the real-time retarded Green's functions can be calculated to obtain final results of these transport coefficients, which in turn are the basis of numerical simulation of the evolution of spinful fluids. According for the spin degree of freedom, one would need to perform the calculation based on the theory of spinor or vector field, which would be a non-trivial extension of that based on a scalar field theory~\cite{Jeon:1992kk}. Then it would be  interesting to see to what extent the results obtained are the same compared to the calculation of ~\cite{Montenegro:2020paq}, which is also based on linear response
theory. On the other hand, it is an efficient way to use transport methods to determine these coefficients if the Green's functions show good behavior during some period of the evolution of the system \cite{Arnold:1997gh}. In that case, we are able to perform simulation within a transport model like \cite{Greif:2014oia,Chen:2018mwr,Chen:2019usj}. The more straightforward way to calculate transport coefficients is to linearize quantum transport equations bypass the path of Kubo formulation. Such a study will be performed in future.

\section{Acknowledgments} 
J.H. is grateful to Guojun Huang for helpful discussions and to Shuzhe Shi, Zhengyu Chen and Ziyue Wang for reading the manuscript and valuable comments and discussions. This work was supported by the NSFC Grant No.11890710 and No.11890712.

\bibliographystyle{apsrev}
\bibliography{KuboW}

\end{document}